\documentclass[lettersize,journal]{IEEEtran}

\usepackage{amsmath,amsfonts}
\usepackage{algorithmic}
\usepackage{algorithm}
\usepackage{array}
\usepackage{color}
\usepackage[caption=false,font=normalsize,labelfont=sf,textfont=sf]{subfig}
\usepackage{textcomp}
\usepackage{stfloats}
\usepackage{url}
\usepackage{verbatim}
\usepackage{graphicx}
\usepackage{cite}
\usepackage{graphicx}
\usepackage{algorithm}
\usepackage{algorithmic}
\usepackage{graphicx} 
\usepackage{float} 
\usepackage{diagbox}
\usepackage{algorithm}
\usepackage{algorithmic}
\usepackage{multirow}
\usepackage{makecell}
\usepackage{amsmath}
\usepackage{caption}
\usepackage[table,xcdraw]{xcolor}
\usepackage{colortbl}
\usepackage{url,array}
\usepackage{amssymb}
\usepackage{hyperref}
\usepackage{booktabs}

\begin{document}

\title{FPO: Fine-grained Preference Optimization Improves Zero-shot Text-to-Speech}

\author{Jixun Yao, Yang Yuguang, Yuan Feng, Yu Pan, Ziqian Ning, Jianhao Ye, \\ Hongbin Zhou, Lei Xie, ~\IEEEmembership{Senior member,~IEEE,}}

\markboth{Journal of \LaTeX\ Class Files,~Vol.~14, No.~8, August~2021}%
{Shell \MakeLowercase{\textit{et al.}}: A Sample Article Using IEEEtran.cls for IEEE Journals}


\maketitle

\begin{abstract}
Integrating reinforcement learning to align generated speech with human preferences has proven effective in improving the robustness of language model-based text-to-speech (TTS) systems. Current approaches primarily rely on preference data annotated at the utterance level. However, frequent issues affecting the listening experience, such as unnatural prosody, mispronunciations, or semantic inconsistencies, often arise only in specific segments of audio, while other segments may be well-generated and require no correction. This mismatch between coarse-grained supervision and fine-grained quality variation limits the effectiveness of preference-based optimization.
In this study, we propose a fine-grained preference optimization approach (FPO) to enhance the robustness of TTS systems. FPO shifts the optimization paradigm from global utterance-level tuning to targeted local refinement, focusing on addressing localized issues in generated samples rather than uniformly optimizing the entire utterance. We begin by analyzing the types of common generation issues and categorizing them into temporal modeling errors and semantic-phonetic alignment errors, which frequently degrade intelligibility and naturalness. To tackle these problems, we introduce a selective training loss strategy that leverages fine-grained labels for each issue type, allowing the model to focus on learning signals where they are most needed.
Experimental results demonstrate that FPO substantially improves the robustness of zero-shot TTS systems by effectively correcting problematic regions in the output. This leads to a significant reduction in the bad case ratio, improved intelligibility, and overall perceptual quality. Moreover, FPO exhibits strong data efficiency, achieving comparable or superior performance to baseline methods while requiring notably fewer training samples.
Audio samples can be found at \url{https://yaoxunji.github.io/fpo/}.
\end{abstract}


\section{Introduction}
Large language models (LLMs) demonstrate remarkable in-context learning abilities in natural language processing (NLP) tasks through scaling model parameters and dataset sizes~\cite{brown2020language,touvron2023llama,grattafiori2024llama3}. This scaling has enabled them to generalize across a wide range of tasks with minimal supervision, often achieving near-human or even superhuman performance in many domains. However, their vast and diverse training data encompasses varied goals, priorities, and content, not all of which are desirable to imitate. Selecting desired responses and behaviors from the model’s broad knowledge and capabilities is crucial for developing safe, high-performing, and controllable AI systems~\cite{christiano2017deep}. A key solution involves aligning LLMs with human preferences through reinforcement learning (RL), effectively calibrating outputs to reduce harmful content risks.  This approach underpins widely-used systems like ChatGPT~\cite{achiam2023gpt} and has recently expanded beyond pure language generation to multimodal tasks, including vision-language understanding~\cite{kelly2024visiongpt}, code generation~\cite{zheng2024opencodeinterpreter}, and speech processing~\cite{yao2025gense}.

Speech, a key modality in generative artificial intelligence, is experiencing rapid advancements inspired by developments in text-based LLMs. Recent text-to-speech (TTS) studies~\cite{wang2023valle,chen2024vall2,du2024cosyvoice} tokenize speech signals into discrete units via neural codec models, aligning them with phoneme or text inputs. These tokens are then processed by a language model to predict subsequent tokens in a next-token prediction manner using phoneme inputs and speech prompts. By scaling up model parameters and training data size, the in-context learning capabilities also emerge in TTS language models, enabling zero-shot generation of high-quality, personalized speech. Mirroring RL’s role in refining text LLMs, emerging interest focuses on applying RL to TTS systems~\cite{zhang2024speechalign,chen2024uno,hu2024rio,tian2024tx_rl}. Typical approaches introduce direct preference optimization (DPO)~\cite{rafailov2024dpo} or Kahneman-Tversky Optimization (KTO)~\cite{ethayarajh2024kto} to align TTS systems, while the key challenge is to collect desirable and relevant preference datasets. By annotating generated speech samples using professional annotators at the utterance-level, these approaches effectively enhance the speech quality and zero-shot capabilities of TTS systems by post-training in an RL manner.

Despite significant advancements in integrating RL into TTS systems, we believe two primary challenges remain. The first challenge is efficiently calibrating pre-trained TTS systems with limited preference data. RL in TTS studies typically requires preference data evaluated by human annotators, which is then used to align TTS systems with human preferences. However, obtaining large-scale human-annotated speech preference data is prohibitively expensive. Compared to annotating text data, annotating speech data is more difficult due to its inclusion of both linguistic content and paralinguistic information.



The second challenge is achieving fine-grained alignment of specific segments in generated samples with human feedback. While issues like poor speaker similarity and degraded audio quality—referred to as \textbf{systemic errors}—have largely been mitigated in TTS systems by scaling up training data, more frequent issues affecting the listening experience often occur in specific segments of the audio. We refer to these as \textbf{segmental errors}, which include mispronunciations, abnormal pauses, repetitions, and other disruptions. Current speech RL approaches typically annotate preference data at the utterance level, failing to capture these fine-grained issues. Moreover, using utterance-level preference data to optimize TTS systems needs to calculate losses of the entire sequence, even when many segments are already well-trained, leading to ineffective gradient updates and inefficient optimization \cite{lin2024rho}. We believe that a more fine-grained approach to optimizing problematic segments would be both more effective and data-efficient.


To address the above challenges, we propose a novel fine-grained preference optimization (FPO) framework for TTS systems—an RL-based approach designed to address segmental errors in generated samples and enhance the robustness of zero-shot TTS systems. FPO employs a fine-grained sampling-annotation pipeline for preference data collection, which plays a crucial role in the RL-based approaches. Pre-trained TTS systems first generate representative audio samples multiple times as the candidate preference data, while the utterance-level preference and dispreference data are annotated from human listening perspectives using these candidate data. For each paired preference data, we further identify the problematic segment and annotate it with fine-grained token-level preference labels, enabling more precise optimization than traditional methods. Furthermore, we analyze the types of issues across different segments from an evaluation perspective and divide the two types of errors in the annotated dispreference data. A selective training loss strategy is proposed to optimize different types of issues during preference optimization. Specifically, FPO processes the entire sequence through the TTS models but computes loss exclusively for the problematic segments, which can effectively resolve fine-grained issues while enhancing training efficiency. Experimental results demonstrate that FPO significantly enhances the robustness of zero-shot TTS systems and exhibits superior data efficiency compared to baseline systems, improving the quality of generated speech across both subjective and objective metrics. 
The main contributions can be summarized as follows:



\begin{itemize}
    \item We propose FPO, an RL-based approach designed to align problematic segments in generated samples, thereby enhancing the robustness of TTS systems. The key novelty of FPO lies in its fine-grained preference optimization, which focuses on optimizing specific problematic segments rather than entire utterances. This approach avoids unnecessary computation on well-learned tokens, resulting in a more data-efficient optimization process.
    \item By analyzing issues in generated samples, FPO optimizes TTS systems by aligning problematic segments through a selective training strategy. This approach offers a new perspective for addressing fine-grained fixes in TTS preference optimization, which resolves bad cases both efficiently and effectively.
    \item Both subjective and objective experiments demonstrate the efficiency and effectiveness of FPO. Notably, FPO significantly reduces the occurrence of bad cases compared to baseline systems and can achieve similar performance with fewer training samples.
\end{itemize}

\section{Related Work}
\subsection{Neural codec based Zero-shot TTS}

Inspired by the success of LLMs in NLP, formatting the TTS task as a next-token prediction problem has gained significant popularity in recent years~\cite{borsos2022audiolm,zhang2023speechgpt}. This formulation allows TTS systems to benefit from autoregressive modeling and in-context learning capabilities that have proven effective in text generation tasks. VALL-E~\cite{wang2023valle}, as a pioneering work in this direction, introduced the first neural codec-based TTS framework, leveraging large-scale speech data to develop strong generalization and prompt-based adaptation abilities. It can synthesize high-quality, personalized speech in a zero-shot manner using only a 3-second enrolled recording as an acoustic prompt, without requiring speaker-specific fine-tuning or additional adaptation steps. Building on this foundation, a series of neural codec-based TTS systems~\cite{kharitonov2023speak,xin2024ralle,peng2024voicecraft} have been developed. These models typically involve a three-stage pipeline: (1) discretizing continuous speech waveforms into semantic and/or acoustic tokens using a neural audio codec~\cite{defossez2022high,zeghidour2021soundstream,zhang2023speechtokenizer,ji2024wavtokenizer}, (2) training a decoder-only language model to predict these tokens conditioned on phoneme sequences and optional prompts, and (3) reconstructing the speech waveform from predicted tokens using the codec’s decoder. This discrete tokenization enables autoregressive modeling and makes the TTS task more analogous to language modeling.

Another prevalent approach introduces self-supervised speech representation models~\cite{hsu2021hubert,chen2022wavlm,chung2021w2v} to extract semantic tokens that capture linguistic content in a speaker-agnostic manner. These semantic tokens serve as intermediate targets for language models, which eases the token prediction task compared to directly modeling raw acoustic tokens\cite{defossez2022high}. To reconstruct the waveform, many systems further employ diffusion-based vocoders or conditional decoders that synthesize speech from the generated semantic tokens, optionally guided by acoustic features or speaker embeddings. These neural codec-based frameworks have demonstrated strong scalability with increasing data and model size, enabling high-fidelity speech synthesis and robust zero-shot generalization to unseen speakers, speaking styles, and acoustic conditions~\cite{lajszczak2024base,du2024cosyvoice}. Despite their impressive performance, these models still face challenges in terms of robustness, particularly under mismatched conditions, and often suffer from localized quality degradations that are not effectively addressed by global training objectives.

\begin{figure*}[ht]
  \centering
  \includegraphics[width=13cm]{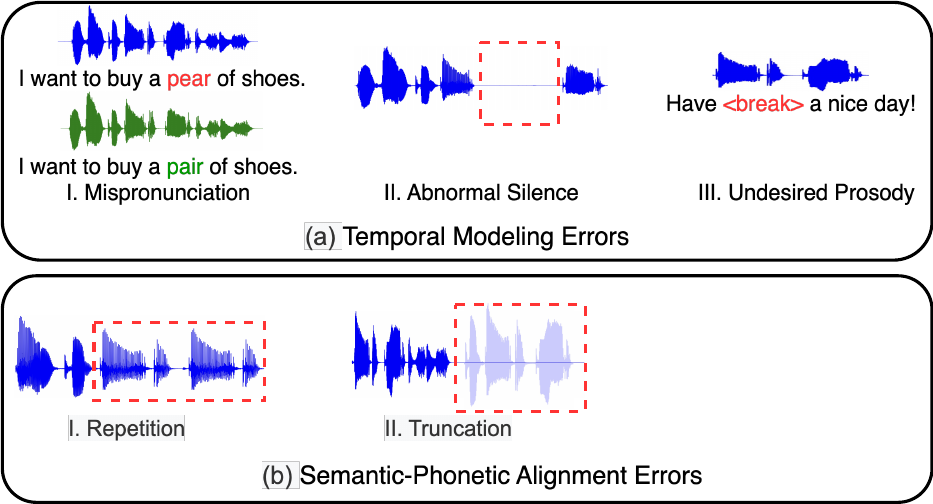}
  \caption{Typical examples of segmental errors can be categorized into two types: temporal modeling errors and semantic-phonetic alignment errors. The problematic segments in the generated samples are highlighted using red boxes or symbols.}
  \label{fig:example}
\end{figure*}

\subsection{Learning from Human Feedback}


Human feedback has been widely utilized in LLMs for NLP tasks~\cite{ouyang2022training,stiennon2020learning}, where RL techniques play a crucial role in aligning model behavior with human preferences. This alignment paradigm has been instrumental in developing AI agents that are helpful, harmless, and honest, paving the way for controllable language models. Traditionally, RL-based preference optimization follows a two-step pipeline. First, a reward model is trained on human-annotated preference data, where annotators compare model outputs and indicate which one is preferred. Then, the language model is fine-tuned using reinforcement learning, typically with Proximal Policy Optimization (PPO)~\cite{schulman2017proximal}, to maximize the expected reward as estimated by the reward model. While effective, this approach is computationally intensive and sensitive to reward model quality and policy instability during RL fine-tuning.

To address these limitations, recent approaches such as Direct Preference Optimization (DPO)~\cite{rafailov2024dpo} propose a more efficient alternative. DPO introduces a closed-form loss function that operates directly on preference pairs, eliminating the need for an explicit reward model or reinforcement learning loop. This formulation simplifies the training pipeline while achieving alignment performance comparable to traditional RLHF methods. Other variants~\cite{gu2025mask}, such as Implicit Preference Optimization (IPO)~\cite{garg2025ipo} and Kahneman-Tversky Optimization (KTO)~\cite{ethayarajh2024kto}, explore related directions to improve preference learning stability and generalization. Although these methods have been extensively applied to NLP tasks, their adoption in speech-related domains remains relatively limited. Recent works begin exploring learning from human preferences in speech synthesis and voice conversion, yet most rely on utterance-level binary labels, which can be insufficient to capture the fine-grained nature of perceptual quality issues in speech.

Learning from human feedback in TTS remains in its infancy compared to its advancements in NLP. SpeechAlign first introduced RLHF in TTS~\cite{zhang2024speechalign}, exploring multiple preference alignment methods using paired preference data. UNO~\cite{chen2024uno} and RIO~\cite{hu2024rio} extended this by optimizing unpaired preference data: UNO accounts for annotation uncertainty in subjective evaluations, while RIO introduces a reverse preference data selection method based on Bayesian principles. Recent studies~\cite{tian2024tx_rl} have conducted thorough empirical evaluations of how preference alignment algorithms enhance LM-based TTS, and Seed-TTS~\cite{anastassiou2024seedtts} has also adopted the RL method during the post-training stage of LM-based TTS. However, current RL-based speech synthesis approaches are limited to utterance-level preference optimization and overlook the potential of fine-grained alignment. While token-level alignment methods (e.g., Mask-DPO~\cite{gu2025maskdpo}, Rho-1~\cite{lin2024rho}) have been explored in NLP, we adapt and extend it to speech-specific issues such as temporal modeling and semantic-phonetic misalignment, which are absent in text-based tasks.

\section{Methodology}
\subsection{Background}
\textbf{Neural codec based zero-shot TTS} comprises two components: a neural codec model and a language model. The codec model tokenizes the speech signal into discrete acoustic tokens $a \in \mathbb{R}^{n \times l}$, where $n$ is the number of quantizer layers in the codec model and $l$ is the temporal length. The tokenized acoustic tokens can later be reconstructed into speech signals by the codec decoder. Given a text transcript $t$, the language model predicts the acoustic tokens using a next-token prediction approach. A widely used codec model employs a single quantizer~\cite{ye2025llasa}, resulting in the extracted acoustic token being a single sequence along the temporal dimension. This setup simplifies the process of directly concat $t$ and $a$ and computing their joint probabilities in an auto-regressive manner:
\begin{equation}
    p(x)=\prod_{i=1}^l p\left(a_l \mid a_1, \ldots, a_{l-1}, t\right),
\end{equation}
it can also be regarded as a speech continuation task, conditioned on both the text and prompt speech during inference.

\textbf{RL with Preference Data.} The language model is first employed to generate paired speech samples ($y_1$,$y_2$) with prompts $x$. Humans then evaluate these pairs and express preferences for one sample, denoted as $y_w \succ y_l \mid x$, where $y_w$ and $y_l$ represent the preferred and dispreferred samples respectively.
The probability of preference data can be modeled using a reward model, which is trained via maximum likelihood estimation. During the reinforcement learning phase, the reward model provides feedback to the language model, which is then optimized by maximizing the rewards.


\subsection{Preference Data Collection}
The selection of preference data plays a critical role in the RL process. We propose a fine-grained sampling-annotation pipeline that begins with annotating pairwise preference data. Specifically, we first sample speech prompts from an unseen speaker pool for each text transcript and use them as input to the TTS system. To promote more diverse zero-shot TTS generation, we adjust sampling hyperparameters during inference and perform multiple inference times for each prompt, constructing a generated sample dataset $\{y_1^b, y_2^b, \ldots, y_k^b\}$, where $b$ is the $b$-th inference process.

After the sampling process, human annotators evaluate the generated samples from each inference process, identifying preferred and dispreferred ones. To mitigate the high human resource demands for annotations, we draw inspiration from previous studies~\cite{lin2024alignslm} and use pre-trained speech perceptual systems to simulate human feedback. We propose a comprehensive scoring method that combines multiple metrics—including intelligibility, quality, similarity, and duration—to determine the final score of each sample. This scoring system can identify preferred samples based on their final scores $s(y)$:

\begin{small}
    \begin{equation}
    s(y) = \lambda_w \cdot w(y)^p + \lambda_m \cdot m(y)^p + \lambda_c \cdot c(y)^p +\lambda_d \cdot \text{Dur}(y)^p,
\end{equation}
\end{small}
where $w(\cdot)$ and $m(\cdot)$ represent the intelligibility evaluation metric and quality evaluation metric, respectively. $c(y)$ measures the speaker similarity and $\text{Dur}(y)$ measures the duration difference between the generated samples and ground truth samples. The weights $\lambda_*$ normalize each metric to ensure they operate at the same amplitude and scale the range of $s(y)$ at [0,1]. Meanwhile, $p$ is used to increase sensitivity to metrics with small changes.
These metrics are computed using pre-trained automatic systems, simulating various aspects of human evaluation. Therefore, the selection of preference data can be described as follows:
\begin{align}
    (y_l, y_w) \in \Big\{
    y_l = \min& \{s(y)\}, \, y_w = \max \{s(y)\}, \notag \\
    &y_w - y_l > \tau
    \Big\},
\end{align}
where $\tau$ is introduced to constrain sufficient differences between the preferred and dispreferred samples. We set $\tau=0.3$ to enhance the reliability of preference data for training. 


\begin{figure*}[ht]
  \centering
  \includegraphics[width=14
  cm]{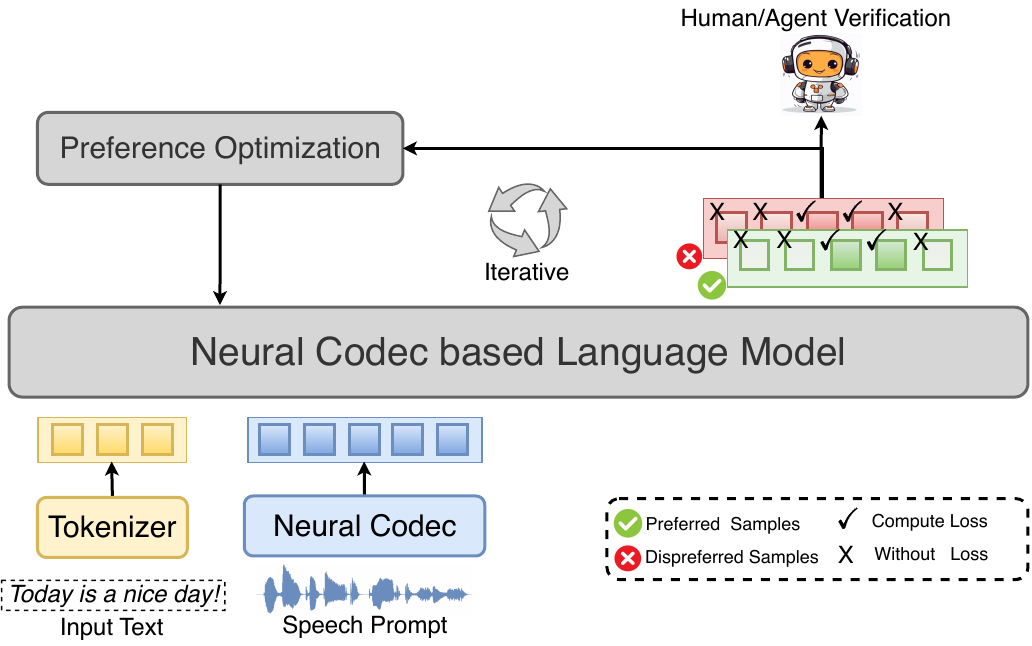}
  \caption{The preference optimization process of our proposed FPO. The language model is optimized using a selective training loss strategy, which focuses loss on the problematic segments while avoiding well-learned tokens.}
  \label{fig:model}
\end{figure*}

\subsection{Fine-grained Annotation}
To achieve fine-grained alignment of TTS systems, we first thoroughly investigate the characteristics of segmental errors and categorize them into two types: temporal modeling errors and semantic-phonetic alignment errors. As illustrated in Figure \ref{fig:example}, temporal modeling errors stem from the model's limitations in capturing the temporal dynamics of speech, leading to issues such as mispronunciation, where abrupt or unstable pronunciation changes occur due to insufficient modeling of pitch continuity or variation trends. Another common problem is abnormal silence, where the model erroneously inserts silent frames or extends silent segments, disrupting the naturalness of generated speech. Additionally, the generated speech may exhibit unnatural prosody (i.g. unnatural pause), as the model may fail to adequately capture local rhythmic patterns, including stress, duration, and speaking rate, resulting in speech that deviates from natural prosody.

For mispronunciations and abnormal silences, we employ Whisper-timestamped\footnote{\url{https://github.com/linto-ai/whisper-timestamped}} and voice activity detection systems to annotate the timestamps of these errors. 
To address unnatural prosody, we leverage the capabilities of text LLMs and professional human annotators to precisely identify affected segments. Specifically, we focus on abnormal pauses and annotate fine-grained prosody errors through a three-step process. First, we use an automatic speech recognition system to transcribe the speech into text. Next, we employ text LLMs to determine optimal pause placements and insert break tags accordingly. Finally, professional annotators evaluate the speech samples using these break tags, identifying preferred and dispreferred prosody patterns. Fine-grained timestamps are then obtained using force-alignment tools based on the text and break tags.
This allows us to obtain timestamps for problematic segments and annotate fine-grained preference labels for temporal modeling errors at the token level.

Semantic-phonetic alignment errors, on the other hand, arise from the mapping between phonetic input and acoustic output. These errors often manifest as repetitions or truncations, where the TTS system either over-relies on its previous outputs or fails to encode essential phonetic information. Alternatively, truncation errors may occur when the language model fails to encode key phonetic information from the input or neglects certain critical elements, leading to synthesized speech that lacks completeness and accuracy. Annotating these error segments relies on automatic speech recognition systems and ground truth transcripts with force alignment tools. 
However, unlike temporal modeling errors, which affect only specific segments, repetition and truncation errors often cause cascading issues after the initial error segment. Therefore, we apply different annotation strategies for each error type. For temporal modeling errors, only the problematic segments are annotated as error segments. In contrast, for semantic-phonetic alignment errors, all segments following the onset of the issue are annotated as error segments. The token-level preference annotation can be described as an indicator function:
\begin{equation}
    I\left(y^i\right)= \begin{cases}1 & \text { if } y^i \text{ in the error segment }  \\ 0 & \text { otherwise }\end{cases},
\end{equation}
where $i$ represents the $i$-th token of preference data.

In the fine-grained annotation process, preference data is first identified, followed by the annotation of fine-grained preference labels by human annotators or automated agents—steps not included in conventional annotation methods. While fine-grained annotation involves additional steps, it requires less data than the traditional approach. Although utterance-level annotation is faster for human annotators, it demands more data to achieve the same level of alignment as fine-grained preference optimization. Therefore, we believe the additional cost of fine-grained annotation is both more acceptable and more efficient than conventional utterance-level annotation.


\subsection{Preference Optimization}
The preference optimization process of our proposed FPO is illustrated in Figure~\ref{fig:model}.
A typical preference optimization process of RL with a reward model $r_\phi(x, y)$ can be formulated as
\begin{equation}
    \max _{\pi_\theta} \mathbb{E}\left[r_\phi(x, y)\right]-\beta \mathbb{D}_{\mathrm{KL}}\left[\pi_\theta(y \mid x) \| \pi_{\mathrm{ref}}(y \mid x)\right],
\end{equation}
where $\beta$ is a parameter controlling the deviation between the TTS model $\pi_\theta$ and the reference model $\pi_{\text{ref}}$. 
However, estimating a reward model is computationally expensive and introduces additional complexity during training. To address this, we follow the DPO approach~\cite{rafailov2024dpo} to directly align the TTS system using preference data, eliminating the need for a reward model. 

Following previous studies~\cite{korbak2022reinforcement,go2023aligning}, the optimal solution to the KL-constrained reward maximization objective in Eq.(5) can be described as follows:
\begin{equation}
    \pi_r(y \mid x)=\frac{1}{Z(x)} \pi_{\text {ref }}(y \mid x) \exp \left(\frac{1}{\beta} r(x, y)\right).
\end{equation}
where $r(x, y)$ is the reward function widely defined as $r(x,y)=r_\phi(x, y)-\beta(\log \pi_\theta(y|x)-\log \pi_{\text{ref}}(y|x))$, 
while $r(x, y)$ in optimization can be expressed in terms of the corresponding optimal policy, the reference policy, and the unknown partition function $Z(\cdot)$. This reward function is formulated as 
\begin{equation}
    r(x, y)=\beta \log \frac{\pi_r(y \mid x)}{\pi_{\mathrm{ref}}(y \mid x)}+\beta \log Z(x).
\end{equation}
Since the Bradley-Terry model depends only on the difference in rewards between two completions, the partition function $Z(\cdot)$ can be canceled, allowing the human preference probability to be expressed solely in terms of the optimal TTS model and the reference model. Finally, to achieve fine-grained optimization, we introduce the indicator function from Eq. (4) to compute selective training loss for both preferred and dispreferred samples:
\begin{align}
\mathcal{L}_{\text{FPO}}= 
- \mathbb{E} 
\Bigg[ & \sum_{i} I(y^i) \cdot \log \sigma \Big( \beta \log \frac{\pi_\theta(y_w^i | x)}{\pi_{\text{ref}}(y_w^i | x)} \nonumber \\
& - \beta \log \frac{\pi_\theta(y_l^i | x)}{\pi_{\text{ref}}(y_l^i | x)} \Big) \Bigg].
\end{align}
This function allows us to fit a token-level implicit reward to optimize the TTS model using preference data while avoiding computational waste on well-trained tokens.

\section{Experimental Setup}

\begin{table*}[ht]
\centering
\caption{Objective evaluation results on WER/CER, SECS, and bad case ratio with different languages. For WER/CER and bad case ratio, lower values indicate better performance, while for SECS, higher values are better. The results are averaged by five inference times and the best in each category are highlighted in bold.
}
\label{tab:comp}
\renewcommand\arraystretch{1.3}
\begin{tabular}{lccccccccc}
\hline
           & \multicolumn{2}{c}{test-zh} & \multicolumn{2}{c}{test-en} & \multicolumn{2}{c}{test-hard} & \multirow{2}{*}{UTMOS $\uparrow$} & \multirow{2}{*}{Bad Case (\%) $\downarrow$} & \multirow{2}{*}{F0 Corr. $\uparrow$}\\ \cline{2-7}
           & CER $\downarrow$          & SECS $\uparrow$        & WER $\downarrow$         & SECS $\uparrow$        & WER $\downarrow$          & SECS $\uparrow$          &                                      \\ \hline
CosyVoice \cite{du2024cosyvoice}  & 3.63\textcolor{gray}{$_{-0.0\%}$}         & 0.723        & 4.29\textcolor{gray}{$_{-0.0\%}$}         & 0.609        & 11.75\textcolor{gray}{$_{-0.0\%}$}         & 0.709       & 3.38$_{\pm0.11}$  & 21\%   & 0.71                                \\
\quad + SDPO \cite{rafailov2024dpo}       & 2.69\textcolor{purple}{$_{-25.9\%}$}         & 0.731        & 3.41\textcolor{purple}{$_{-20.5\%}$}         & 0.645        & 9.84\textcolor{purple}{$_{-16.3\%}$}          & 0.716      & 3.44$_{\pm0.10}$   & 17\%                 & 0.70                \\
\quad + UNO \cite{chen2024uno}       & 2.57\textcolor{purple}{$_{-29.2\%}$}         & 0.735        & 3.06\textcolor{purple}{$_{-28.7\%}$}         & 0.667        & 9.06\textcolor{purple}{$_{-22.9\%}$}          & 0.722      & \textbf{3.67}$_{\pm0.11}$   & 15\%           & 0.75                      \\
\quad + RIO \cite{hu2024rio}        & 2.21\textcolor{purple}{$_{-39.1\%}$}         & \textbf{0.744}        & 2.89\textcolor{purple}{$_{-32.6\%}$}         & \textbf{0.681}        & 9.12\textcolor{purple}{$_{-22.4\%}$}          & \textbf{0.727}     & 3.64$_{\pm0.10}$    & 12\%   & 0.73                              \\
\quad + Mask-DPO \cite{gu2025maskdpo}        & 2.61\textcolor{purple}{$_{-28.1\%}$}         & 0.734        & 3.27\textcolor{purple}{$_{-23.8\%}$}         & 0.652        & 9.52\textcolor{purple}{$_{-19.0\%}$}          & 0.715     & 3.48$_{\pm0.10}$    & 14\%   & 0.65                              \\
\quad + FPO (Ours)       & \textbf{1.83}\textcolor{purple}{$_{-49.6\%}$}         & 0.728        & \textbf{2.77}\textcolor{purple}{$_{-35.4\%}$}         & 0.614        & \textbf{7.98}\textcolor{purple}{$_{-32.1\%}$}          & 0.708      & 3.47$_{\pm0.09}$   & \textbf{9}\%    & \textbf{0.78}                              \\ \hline
CosyVoice2 \cite{du2024cosyvoice2} & 1.45\textcolor{gray}{$_{-0.0\%}$}         & 0.748        & 2.57\textcolor{gray}{$_{-0.0\%}$}         & 0.652        & 8.08\textcolor{gray}{$_{-0.0\%}$}          & 0.730      & 3.49$_{\pm0.11}$   & 14\%       & 0.75                          \\
\quad + SDPO \cite{rafailov2024dpo}      & 1.43\textcolor{purple}{$_{-1.4\%}$}         & 0.745        & 2.71\textcolor{teal}{$_{+5.4\%}$}         & 0.673        & 8.32\textcolor{teal}{$_{+3.0\%}$}          & 0.728       & 3.59$_{\pm0.08}$  & 15\%            & 0.73                     \\
\quad + UNO \cite{chen2024uno}       & 1.37\textcolor{purple}{$_{-5.5\%}$}         & \textbf{0.759}        & 2.43\textcolor{purple}{$_{-5.4\%}$}         & 0.669        & 7.78\textcolor{purple}{$_{-3.7\%}$}          & 0.736        & 3.61$_{\pm0.11}$ & 12\%       & 0.75                          \\
\quad + RIO \cite{hu2024rio}       & 1.40\textcolor{purple}{$_{-3.4\%}$}         & 0.751        & 2.47\textcolor{purple}{$_{-3.9\%}$}         & \textbf{0.675}        & 7.94\textcolor{purple}{$_{-1.7\%}$}          & \textbf{0.737}     & \textbf{3.73}$_{\pm0.12}$    & 9\%      & 0.77                            \\
\quad + Mask-DPO \cite{gu2025maskdpo}        & 1.42\textcolor{purple}{$_{-2.1\%}$}         & 0.754        & 2.51\textcolor{purple}{$_{-2.3\%}$}         & 0.671        & 7.92\textcolor{purple}{$_{-2.0\%}$}          & 0.733     & 3.64$_{\pm0.10}$    & 13\%                & 0.67                 \\
\quad + FPO (Ours)       & \textbf{1.32}\textcolor{purple}{$_{-9.0\%}$}         & 0.747        & \textbf{2.24}\textcolor{purple}{$_{-12.8\%}$}         & 0.654        & \textbf{6.92}\textcolor{purple}{$_{-14.4\%}$}          & 0.731      & 3.56$_{\pm0.10}$   & \textbf{8}\%         & \textbf{0.81}                         \\ \hline
\text{Llasa \cite{ye2025llasa}} & 1.89\textcolor{gray}{$_{-0.0\%}$} & 0.669 & 3.22\textcolor{gray}{$_{-0.0\%}$} & 0.572 & 12.13\textcolor{gray}{$_{-0.0\%}$} & 0.638 & 3.54$_{\pm0.12}$ & 25\% & 0.73 \\ \quad + \text{SDPO \cite{rafailov2024dpo}} & 1.73\textcolor{purple}{$_{-8.5\%}$} & 0.685 & 2.83\textcolor{purple}{$_{-12.1\%}$} & 0.599 & 11.23\textcolor{purple}{$_{-7.4\%}$} & 0.633 & 3.67$_{\pm0.11}$ & 22\% & 0.75 \\ \quad + \text{UNO \cite{chen2024uno}} & 1.64\textcolor{purple}{$_{-13.2\%}$} & 0.701 & 2.74\textcolor{purple}{$_{-14.9\%}$} & 0.623 & 10.84\textcolor{purple}{$_{-10.6\%}$} & 0.663 & 3.65$_{\pm0.13}$ & 18\% & 0.74 \\ \quad + \text{RIO \cite{hu2024rio}} & 1.58\textcolor{purple}{$_{-16.4\%}$} & \textbf{0.709} & 2.79\textcolor{purple}{$_{-13.4\%}$} & \textbf{0.625} & 9.45\textcolor{purple}{$_{-22.1\%}$} & \textbf{0.671} & \textbf{3.75}$_{\pm0.11}$ & 15\% & 0.78 \\ \quad + \text{Mask-DPO \cite{gu2025maskdpo}} & 1.71\textcolor{purple}{$_{-9.5\%}$} & 0.694 & 2.84\textcolor{purple}{$_{-11.8\%}$} & 0.591 & 9.92\textcolor{purple}{$_{-18.2\%}$} & 0.637 & 3.64$_{\pm0.10}$ & 19\% & 0.64 \\ \quad + \text{FPO (Ours)} & \textbf{1.47}\textcolor{purple}{$_{-22.2\%}$} & 0.672 & \textbf{2.46}\textcolor{purple}{$_{-23.6\%}$} & 0.571 & \textbf{8.42}\textcolor{purple}{$_{-30.6\%}$} & 0.640 & 3.58$_{\pm0.12}$ & \textbf{11\%} & \textbf{0.79}\\ \hline
\end{tabular}
\end{table*}

\subsection{Model Configuration}
We select the open-source CosyVoice~\cite{du2024cosyvoice}, CosyVoice2~\cite{du2024cosyvoice2} and Llasa~\cite{ye2025llasa} as our backbone models. These systems are trained on large-scale datasets and demonstrate impressive performance. 
CosyVoice series consists of a text encoder, a speech tokenizer, a speech language model and a conditional flow matching model. Specifically, the text encoder aligns the semantic spaces of text and speech tokens, while the speech tokenizer extracts semantic tokens. The speech language model then predicts speech tokens based on text inputs and an acoustic prompt. A conditional flow matching model converts the predicted speech tokens into a mel-spectrogram, which is finally reconstructed into speech signals by a pre-trained vocoder. In contrast, Llasa bypasses flow matching and the vocoder, directly reconstructing the predicted speech tokens into waveforms using a codec decoder.
Our study focuses on optimizing the speech language model using fine-grained preference data.  For model training, we use a learning rate of 2e-6 and a batch size of 64, optimized with the AdamW optimizer~\cite{loshchilov2017adamw}. The model is trained for only two epochs.

\subsection{Dataset}
To construct the training data for preference optimization, we use the LibriTTS~\cite{zen2019libritts} and AISHELL-3~\cite{shi2020aishell3} corpora, which are already included in the base models’ training. This setup eliminates the influence of introducing additional high-quality data. Following a prior study~\cite{hu2024rio}, we sample a pool of speech prompts containing 100 individual speakers and a separate pool of target transcripts to generate preference samples using the pre-trained models. With the proposed fine-grained sampling–annotation pipeline, we then select sample pairs with substantial differences to build the training set, resulting in 1,000 utterances for preference optimization.

For evaluation, we use the Seed-TTS eval dataset \cite{anastassiou2024seedtts} to evaluate model performance after preference optimization. This test set includes samples extracted from publicly available English (EN) and Mandarin (ZH) corpora and is designed to measure performance across various objective metrics. Each sample in the English and Mandarin test sets consists of one reference utterance and one target utterance spoken by the same speaker. The optimized model generates speech for the target text using the reference speech as an audio prompt.

\subsection{Baseline Systems}
In addition to using the CosyVoice series as base models, we reproduce the following strong RL-based speech approaches based on CosyVoice to serve as powerful baseline systems for comprehensive comparison:

\begin{itemize}
    \item Speech-DPO~\cite{rafailov2024dpo}: 
    We adapt the popular DPO algorithm to align the base model with human feedback, optimizing preference data at the utterance level, similar to the approach described in \cite{zhang2024speechalign}. However, instead of using ground-truth and generated samples as pairs, we use both generated samples as the preference and the other as the dispreference. We refer to this baseline as SDPO.
    \item UNO~\cite{chen2024uno}: A recent study proposes an RL framework tailored for zero-shot TTS models, which integrates human feedback into the TTS learning objective and specifically considers the uncertainty arising from inherent variability in subjective human perception and evaluation. The preference optimization maximizes the utility of these samples in an uncertainty-aware manner.
    \item RIO~\cite{hu2024rio}: An RL-based preference optimization approach tailored to the zero-shot TTS system with neural codec modelling. It introduces a reverse inference optimization method based on the Bayesian principle to select exemplars from speech samples generated by the TTS system, which is further employed to generate the original prompt samples. By using the reverse inference method, it can effectively select preference data to enhance the capacity of an advanced zero-shot TTS system.
    \item Mask-DPO~\cite{gu2025maskdpo}: A fine-grained factuality alignment method based on DPO that incorporating sentence-level factuality as mask signals and only learns from factually correct sentences in the preferred samples and prevents the penalty on factual contents in the not preferred samples, which resolves the ambiguity in the preference learning.

\end{itemize}

\subsection{Evaluation Metrics}
\textbf{Subjective Metrics}: We employ the naturalness mean opinion score (NMOS) to evaluate the naturalness of the generated samples. We invited 15 participants to listen to the audio samples and rate the naturalness on a 5-point scale, where 1 indicates very unnatural and 5 indicates completely natural. 
We apply a t-test to assess the statistical significance between systems with 95\% confidence intervals for the mean opinion scores to better reflect the variability of listener ratings.
Furthermore, we conduct an ABX test in which participants listen to two generated speech samples from different models based on the same input and select the more natural-sounding sample. If the samples are too similar to distinguish, participants are instructed to indicate a tie.

\textbf{Objective Metrics}: For objective evaluation, we use the speaker embedding cosine similarity (SECS), calculated using pre-trained WavLM-large fine-tuned automatic speaker verification models\footnote{\url{https://github.com/microsoft/UniSpeech/tree/main/downstreams/speaker_verification}}, to evaluate speaker similarity. 
For intelligibility evaluation, we employ the word error rate (WER) metric for English samples and the character error rate (CER) metric for Mandarin samples. English samples are transcribed using the Whisper-Large-v3 model \cite{radford2022whisper}\footnote{\url{https://huggingface.co/openai/whisper-large-v3}}, while Mandarin samples are transcribed using the Paraformer model \cite{gao2022paraformer}\footnote{\url{https://huggingface.co/funasr/paraformer-zh}}. 
Additionally, speech quality is further measured using a predicted mean opinion score (UTMOS), computed by a neural network model\footnote{\url{https://github.com/sarulab-speech/UTMOS22}}.
Meanwhile, following previous studies~\cite{hu2024rio}, we introduce the bad case ratio to evaluate the robustness of different baseline systems. Samples are classified as bad cases if their UTMOS is below 3, or if their WER exceeds 3\% on the test-zh and test-en sets, or 6\% on the test-hard set. 
Finally, we use the Pearson correlation (F0 Corr.) between the generated and ground-truth pitch sequences as a metric to evaluate prosody performance.

\section{Experimental Results}
\subsection{Objective Evaluation}

We first compare our proposed FPO with baseline systems using objective metrics, results are shown in Table\ref{tab:comp}. For WER and CER, FPO consistently achieves the most significant improvements in speech intelligibility across all backbones. With the CosyVoice model, FPO reduces CER by 49.6\% (from 3.63 to 1.83) and WER by 35.4\% (from 4.29 to 2.77), outperforming all baseline systems, including RIO ($-$39.1\% CER, $-$32.6\% WER) and Mask-DPO ($-$28.1\% CER, $-$23.8\% WER). Similarly, when built upon the stronger CosyVoice2 backbone, FPO further achieves the lowest CER (1.32) and WER (2.24 and 6.92 on the test-en and test-hard sets, respectively), surpassing other optimization methods such as UNO and SDPO. Notably, FPO also maintains desired SECS and UTMOS scores, indicating improved naturalness without compromising perceptual quality.Beyond CosyVoice-based models, we additionally evaluate FPO on Llasa, a more recent TTS backbone. FPO again demonstrates substantial intelligibility gains, reducing CER by 22.2\% (from 1.89 to 1.47) and WER by 23.6\% (from 3.22 to 2.46), while outperforming all other baselines, including Mask-DPO and RIO.

In terms of robustness, FPO achieves the lowest bad case ratios across all settings—reducing them from 21\% to 9\% on CosyVoice, from 14\% to 8\% on CosyVoice2, and from 25\% to 11\% on Llasa, indicating a strong capability to suppress segmental degradation such as mispronunciations and unnatural repetitions. These consistent improvements across diverse architectures validate the generalizability and effectiveness of our fine-grained preference optimization strategy in enhancing both the reliability and perceptual consistency of speech synthesis systems.


While FPO demonstrates only marginal improvements over the base model in metrics like SECS and UTMOS—with RIO and UNO outperforming it in these aspects—this discrepancy stems from the distinct optimization priorities of the approaches. FPO’s core design focuses on resolving localized segmental errors through token-level selective training, whereas baseline systems like RIO and UNO prioritize utterance-level alignment for global attributes such as speaker similarity and overall quality. The base model already achieves strong performance in speaker similarity and speech quality due to its large-scale pre-training, leaving limited room for improvement in these dimensions. Consequently, the gains from RIO and UNO—though statistically significant—represent incremental refinements rather than transformative advancements. In contrast, FPO’s targeted optimization reduces critical segmental errors by 40\% in CER/WER and slashes the bad case ratio by a large margin, directly addressing the most frequent and perceptually salient issues in human evaluations. This trade-off aligns with practical priorities in real-world TTS deployment, where eliminating catastrophic local errors often outweighs minor improvements in global attributes. 


\begin{table}[ht]
\centering
\renewcommand\arraystretch{1.2}
\caption{Subjective NMOS results for the generated speech between baseline systems and FPO with 95\% confidence interval.}
\label{tab:mos}
\resizebox{0.9\linewidth}{!}{
\begin{tabular}{lccc}
\hline
         & CosyVoice & CosyVoice2 & Llasa \\ \hline
Base     & 3.64$_{\pm0.11}$      & 3.81$_{\pm0.07}$       & 3.74$_{\pm0.12}$  \\ \hline
SDPO     & 3.68$_{\pm0.09}$      & 3.83$_{\pm0.11}$       & 3.80$_{\pm0.09}$  \\
Mask-DPO & 3.60$_{\pm0.12}$      & 3.79$_{\pm0.09}$       & 3.71$_{\pm0.11}$  \\
UNO      & 3.69$_{\pm0.11}$      & 3.72$_{\pm0.10}$       & 3.75$_{\pm0.08}$  \\
RIO      & 3.76$_{\pm0.08}$      & 3.87$_{\pm0.08}$       & 3.82$_{\pm0.09}$  \\
FPO      & \textbf{3.83$_{\pm0.09}$}      & \textbf{3.91$_{\pm0.08}$}       & \textbf{3.93$_{\pm0.10}$}  \\ \hline
\end{tabular}}
\end{table}

\begin{figure}[ht]
  \centering
  \includegraphics[width=8cm]{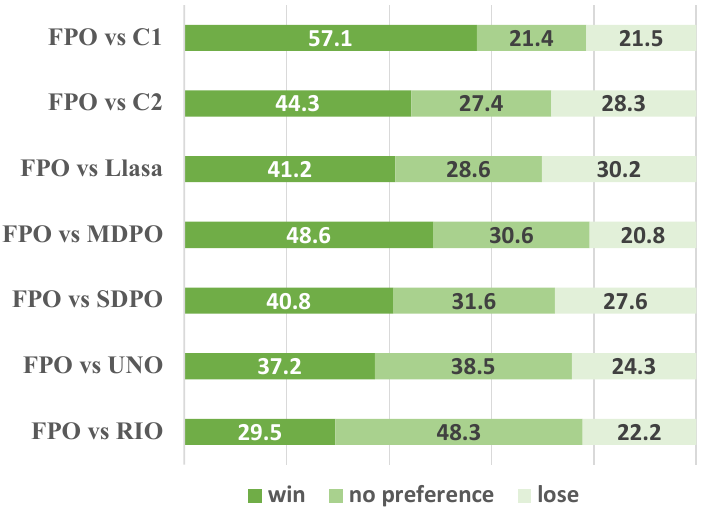}
  \caption{Results of ABX preference test, while "C1" and "C2" denote CosyVoice and CosyVoice2 model, respectively. MDPO denotes Mask-DPO method.}
  \label{fig:abx}
\end{figure}

\begin{figure*}[ht]
  \centering
  \includegraphics[width=16cm]{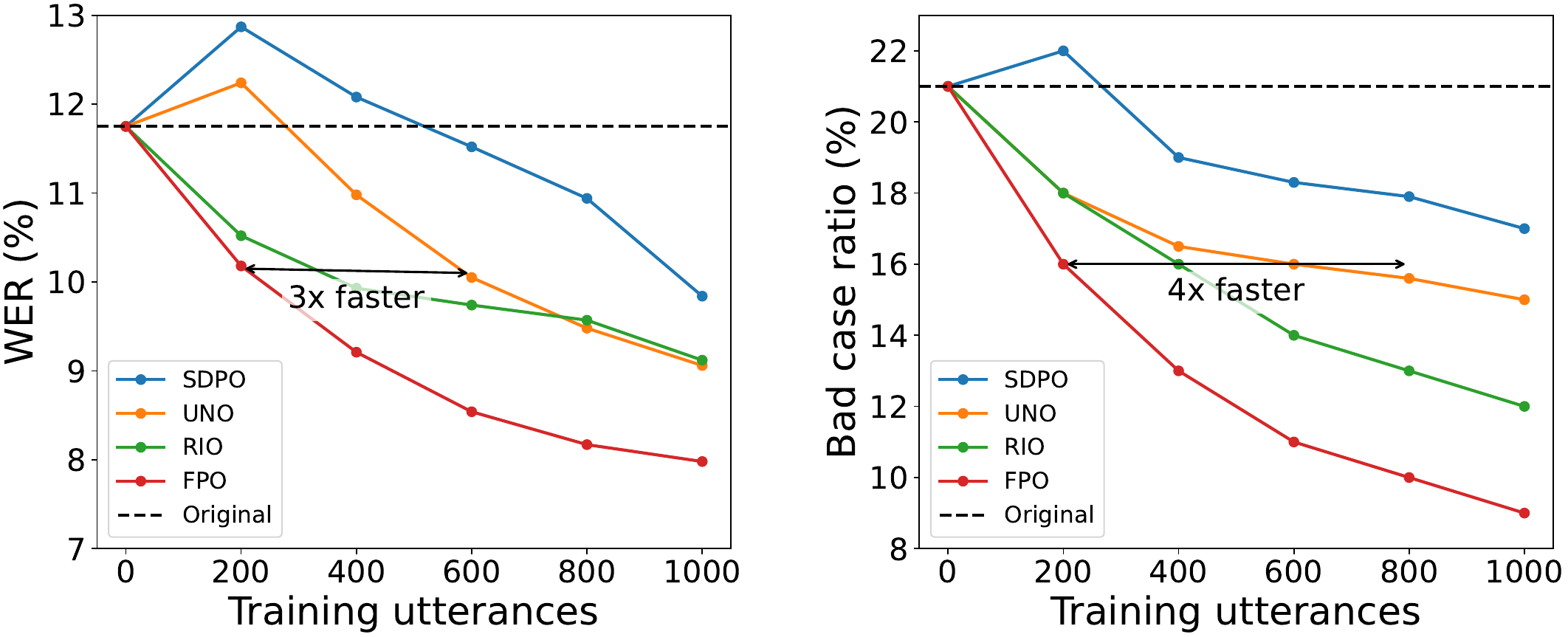}
  \caption{Results of FPO and baseline systems using different scales of training data.}
  \label{fig:efficient}
\end{figure*}

\subsection{Subjective Evaluation}
To further evaluate the effectiveness of FPO, we conduct subjective evaluations on the naturalness of the generated samples. 
We can find that our proposed FPO consistently achieves the highest NMOS scores among all baseline RLHF alignment approaches, as shown in Table \ref{tab:mos}. While several baselines (e.g., RIO and SDPO) exhibit modest increases in naturalness, FPO provides the most substantial and stable enhancement, improving NMOS by +0.19, +0.10, and +0.19 points on CosyVoice, CosyVoice2, and Llasa, respectively. To further investigate the source of these perceptual gains, we combine the objective Pearson correlation results with the subjective evaluations and find that FPO outperforms all baseline systems on both metrics. This indicates that the improved naturalness is not merely due to better intelligibility or fewer segment-level errors. Instead, FPO more effectively aligns prosodic structure, directly contributing to the enhanced perceptual quality.
These findings highlight FPO’s superior performance in generating more natural and consistent speech.


In addition to NMOS, we conduct ABX preference tests to further validate the performance gains of our proposed FPO compared to baseline systems. The ABX tests include both CosyVoice and CosyVoice2 models, while all other baselines are trained based on CosyVoice2. As shown in Figure \ref{fig:abx}, FPO is preferred over CosyVoice in 57.1\% of cases and ties in 21.4\%. Compared to CosyVoice2, FPO achieves a 44.3\% win rate, again significantly outperforming the base model in terms of listener preference. Notably, FPO also outperform SDPO by a substantial margin (40.8\% vs. 27.6\%, with 31.6\% ties), demonstrating the effectiveness of fine-grained optimization over full-sequence optimization. These ABX test results further emphasize the advantages of FPO in producing more preferred and natural speech.

On the other hand, FPO achieves a higher preference score than UNO and RIO in the ABX test. While UNO and RIO focus on enhancing overall attributes such as timbre and quality, FPO specifically targets problematic segments, reducing abrupt errors and improving local consistency. The higher preference scores in ABX tests suggest that fine-grained optimization of local segments better aligns with human perceptual priorities.

These results indicate that while utterance-level optimization methods effectively align the base model to some extent, FPO directly addresses critical points in the human listening experience. This leads to higher practical utility, even when objective metric scores are comparable. The consistency between subjective and objective evaluations further validates the effectiveness of the proposed FPO.

\subsection{Analysis on Data Efficient}

To further evaluate the data efficiency of our proposed FPO, we compare its performance against baseline systems using small-scale English training data, with intelligibility evaluated on the test-hard set using WER as metric and average bad case ratio across three test sets as another metric.
As shown in Figure \ref{fig:efficient}, FPO achieves comparable WER performance to the UNO system using only 200 training utterances, whereas the baselines require 600 utterances. This demonstrates that FPO is up to three times more data-efficient than baseline systems. Furthermore, FPO consistently outperforms all baselines in terms of bad case ratio across various data scales and achieves an approximate 4x speedup compared to UNO.


\begin{table}[ht]
\centering
\renewcommand\arraystretch{1.2}
\caption{Comparison results between FPO and SDPO on several larger-scale datasets. "Num" represents the number of utterances used for training.}
\label{tab:data_comp}
\resizebox{1.0\linewidth}{!}{
\begin{tabular}{lcccc}
\hline
           & Num (\#) & CER & WER & Bad case ratio \\  \hline 
CosyVoice2                   & -                         & 1.45         & 5.33                    & 14\%            \\ \hline
\multirow{3}{*}{\quad+ SDPO} & 1000                             & 1.43         & 5.52         & 15\%                         \\
                            & 2000                      & 1.40         & 5.19                      & 12\%            \\
                            & 5000                      & 1.31         & 3.94                       & 7\%             \\ \hline
\multirow{3}{*}{\quad+ FPO}        & 1000                      & 1.32         & 4.58                       & 8\%             \\ 
                            & 2000                      & 1.29         & 3.76                       & 5\%             \\
                            & 5000                      & 1.21         & 3.23                       & 3\%             \\ \hline
\end{tabular}}
\end{table}

Furthermore, we scale up the preference data size to compare the data scalability between the SDPO approach and our proposed FPO. As shown in Table \ref{tab:data_comp}, with just 1,000 training utterances, FPO reduces the CER of the original CosyVoice2 model from 1.45 to 1.32 and the WER from 5.33 to 4.58 for Mandarin and English evaluation datasets, respectively. This performance outperforms even SDPO trained on 5,000 utterances (CER: 1.32 vs. 1.31; WER: 4.58 vs. 3.94), highlighting FPO’s ability to maximize efficiency of preference optimization with limited data. Notably, FPO reduces the bad case ratio from 14\% to 8\% using only 1,000 samples, whereas SDPO requires 5,000 samples to achieve a similar performance of about 7\%. These results further demonstrate FPO’s effectiveness in mitigating critical errors while maintaining high data efficiency.

FPO’s advantage stems from its selective training loss strategy, which precisely directs optimization toward error-prone segments identified through fine-grained annotations of temporal modeling and semantic-phonetic misalignments. By identifying problematic tokens and excluding well-learned regions from gradient updates, FPO minimizes computational redundancy while preventing over-optimization of already proficient segments. This targeted approach not only reduces noise in parameter updates but also accelerates convergence by prioritizing problematic segments. In contrast, utterance-level methods like SDPO apply uniform optimization across entire sequences, inadvertently diluting the impact of critical segments by averaging gradients from both optimal and suboptimal regions. This indiscriminate updating forces such methods to rely on larger datasets (4–5× more samples in ablation studies) to statistically filter irrelevant signals and achieve comparable error reduction. Furthermore, the utterance-level optimization strategy risks destabilizing well-trained segments through unnecessary perturbations, whereas FPO’s surgical precision preserves existing capabilities while efficiently resolving localized failures.

\begin{table*}[h]
\renewcommand\arraystretch{1.2}
\centering
\caption{Comparison between our proposed FPO and other preference alignment approaches.}
\begin{tabular}{lccccc}
\hline
           & SDPO                                            & UNO                                                    & RIO                                               & Mask-DPO                                 & FPO                                                       \\ \hline
Grain      & Utterance                                       & Utterance                                              & Utterance                                         & Sentence                                 & Token                                                     \\
Goal       & \multicolumn{1}{l}{Overall quality} & \multicolumn{1}{l}{Uncertainty perceptual} & \multicolumn{1}{l}{Robustness enhancement} & \multicolumn{1}{l}{Factuality alignment} & \multicolumn{1}{l}{Targeted correction} \\
Annotation & Global preference                               & Global preference                                      & Global preference                                 & Global preference                        & Token-level preference                                    \\
Reward     & Binary                                          & Binary                                                 & Binary                                            & Binary                                   & Fine-grained                                              \\ \hline
\end{tabular}
\end{table*}

\section{Limitations and Discussion}
One of the main limitations of this study lies in the construction of the preference training dataset. The proposed framework relies on fine-grained manual annotation to ensure the quality and consistency. This process is both time-consuming and labor-intensive, which may limit the scalability of the approach to larger or more diverse datasets. Although our proposed approach is data efficient and need less annotation data compared with previous works, the fine-grained annotation at each sample is more difficult than utterance level annotation. 
Furthermore, our proposed approach is not inherently restricted to autoregressive TTS models. Beyond the evaluated autoregressive backbones, the framework can be naturally extended to discrete non-autoregressive models. In these cases, the underlying mechanism of preference-based optimization remains directly applicable without substantial modification. For TTS paradigms that rely on continuous representations, the core idea of our method is still conceptually applicable. However, applying the method in this setting introduces additional challenges. In particular, constructing sufficiently diverse and well-controlled paired samples from continuous representations for effective reinforcement learning becomes non-trivial. Meanwhile, practical challenges may arise due to differences in training objectives and generation mechanisms, which we leave for future work.


\section{Conclusion}

In this study, we propose FPO, a novel RL framework that introduces a new paradigm for enhancing zero-shot TTS robustness through localized error correction. By systematically categorizing generation errors into temporal modeling discrepancies and semantic-phonetic misalignments, and applying token-level preference optimization, FPO adopts a segment-focused strategy that significantly outperforms baseline systems. This validates our hypothesis that targeted intervention on error-prone segments leads to more efficient and effective error correction. FPO reduces the bad case ratio by 12\% compared to base models through fine-grained perceptual alignment. Additionally, it achieves comparable performance using fewer training samples than baseline systems, demonstrating strong data efficiency. Experimental results confirm that FPO substantially improves the robustness of the base TTS system, while subjective evaluations further support FPO’s superiority in naturalness and alignment with human perceptual preferences.

\bibliographystyle{IEEEtran}
\bibliography{mybib}

\end{document}